\documentclass{article}
\usepackage{spconf,amsmath,amssymb,graphicx,subcaption}

\usepackage[LGR,T1]{fontenc}
\usepackage[colorinlistoftodos,prependcaption]{todonotes}

\usepackage{moredefs} 
\usepackage{color}

\DeclareSymbolFont{upgreek}{LGR}{cmr}{m}{n}
\DeclareMathSymbol{\upsigma}{\mathord}{upgreek}{`s}

\DeclareMathOperator{\decoder}{DecoderRNN}
\DeclareMathOperator{\outpt}{Output}

\title{A COMPARISON OF END-TO-END MODELS FOR LONG-FORM SPEECH RECOGNITION}
\name{\begin{tabular}{c}Chung-Cheng Chiu, Wei Han, Yu Zhang, Ruoming Pang, Sergey Kishchenko, Patrick Nguyen$^{1}$\thanks{$^{1}$Work conducted while the author was at Google},\\Arun Narayanan, Hank Liao, Shuyuan Zhang, Anjuli Kannan, Rohit Prabhavalkar, Zhifeng Chen,\\Tara Sainath, Yonghui Wu\end{tabular}}
\address{Google Inc, Grab Technologies$^{1}$}
\begin{document}
%
\maketitle
\begin{abstract}
End-to-end automatic speech recognition (ASR) models, including both attention-based models and the recurrent neural network transducer (RNN-T), have shown superior performance compared to conventional systems \cite{Chiu2018state,He2019}.  However, previous studies have focused primarily on short utterances that typically last for just a few seconds or, at most, a few tens of seconds. Whether such architectures are practical on long utterances that last from minutes to hours remains an open question. In this paper, we both investigate and improve the performance of end-to-end models on long-form transcription. We first present an empirical comparison of different end-to-end models on a real world long-form task and demonstrate that the RNN-T model is much more robust than attention-based systems in this regime.  We next explore two improvements to attention-based systems that significantly improve its performance: restricting the attention to be monotonic, and applying a novel decoding algorithm that breaks long utterances into shorter overlapping segments.  Combining these two improvements, we show that attention-based end-to-end models can be very competitive to RNN-T on long-form speech recognition.
\end{abstract}
\section{INTRODUCTION}
\label{sec:intro}
End-to-end models have become a popular choice for speech recognition, thanks to both the simplicity of building them and their superior performance over conventional systems \cite{Graves12,Graves2013,Graves14,Chan15,Jan15,Bahdanau16,Battenberg17,RohitSeq17,Yu17,Soltau2017,Chiu2018state,He2019}.  In contrast to conventional systems, which are comprised of separate acoustic, pronunciation, and language modeling components, end-to-end approaches formulate the speech recognition problem directly as a mapping from utterances to transcripts, which greatly simplifies the training and decoding processes.  Popular end-to-end models fall into three broad classes: 1) those that are based on the connectionist temporal classification (CTC) \cite{Graves06} criteria, 2) those that are based on the RNN-T criteria, and 3) those that make use of an attention mechanism.

While recent studies have shown that end-to-end models are very competitive with conventional systems, they have focused mainly on short utterances, which last from a few seconds to a few tens of seconds at most.  Few works have investigated long-form transcription, a capability that is fundamental to applications like continuous transcription of meetings, presentations, or lectures.  In the limited literature on this topic that we are aware of \cite{Soltau2017}, the authors show that end-to-end CTC models can generalize well on long utterances, but it remains unanswered whether RNN-T and the attention-based models can provide the same robustness.  \cite{Jan15} evaluates attention-based models on long-form audio by concatenating multiple short utterances into long utterances, yet it is unclear whether observations on a synthetic data will generalize to real world cases.

In this work, we both evaluate and improve the performance of end-to-end models on long-form audio.  Focusing on RNN-T and attention-based models, we first evaluate popular end-to-end models on a long-form ASR task: the transcription of Youtube videos \cite{Soltau2017}. The Youtube dataset contains a mix of long form audio that closely reflect both common real life use cases (conversations, lectures, TV shows, etc.) and a broad variety of domains (education, sports, etc.).  The training data is automatically generated via the confidence island method as detailed in \cite{Liao2013} and contain utterances with several seconds long, while the test set is human-transcribed and have utterances with a few minutes long.  Comparison on this task shows that standard soft-attention-based models generalize poorly to long audio.

Next, we incorporate two mechanisms in order to improve the generalization of attention-based models to long utterances.  The first mechanism is a monotonicity constraint in the attention model, exploiting the observation that in ASR, the target sequence (transcript) and source sequence (acoustic signal) are monotonically aligned. We explore a few different mechanisms to enforce monotonicity, including the monotonic attention model \cite{Raffel2017}, the monotonic chunkwise attention model \cite{Chiu2018}, the monotonic infinite lookback attention model \cite{Arivazhagan2019}, and the GMM-based monotonic attention model \cite{Graves2013generating,tjandra2017local}.  Our results show that enforcing a monotonicity constraint does improve the generalization of attention-based models to long utterances, but is still not sufficient to fully solve the problem.  Thus we also incorporate a novel decoding algorithm that breaks long utterances into overlapping segments.  We show that the combination of these two mechanisms enables attention-based models to match the performance of RNN-T on a long-form task.

The organization of the rest of the paper is as follows: sections~\ref{sec:models} describes the attention models,  section~\ref{sec:rnnt} describes the RNN-T model evaluated in this work, section~\ref{sec:overlap} gives a brief intro about the decoding strategy that helps the model to be robust to long-form utterances. In section~\ref{sec:exp} we show our evaluation results, and concludes the observations in section~\ref{ssec:conclusions}.


\section{ATTENTION-BASED MODELS}
\label{sec:models}
\begin{figure*}[t]
    \centering
    \begin{subfigure}[b]{0.33\textwidth}
        \includegraphics[width=\textwidth]{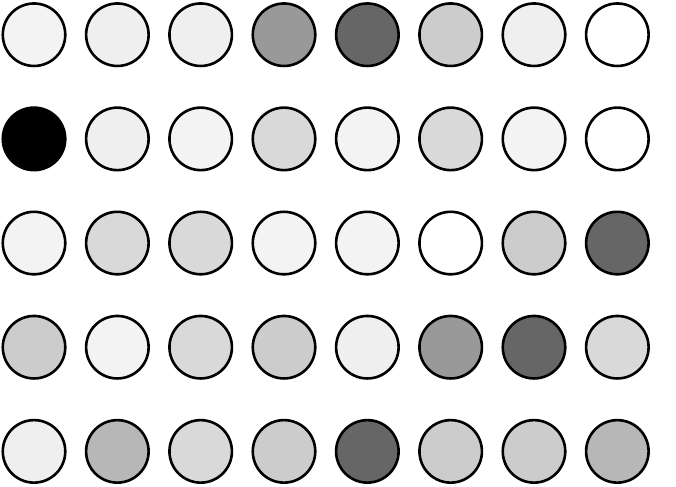}
        \caption{Soft attention.}
        \label{fig:softmax}
    \end{subfigure}
    \begin{subfigure}[b]{0.33\textwidth}
        \includegraphics[width=\textwidth]{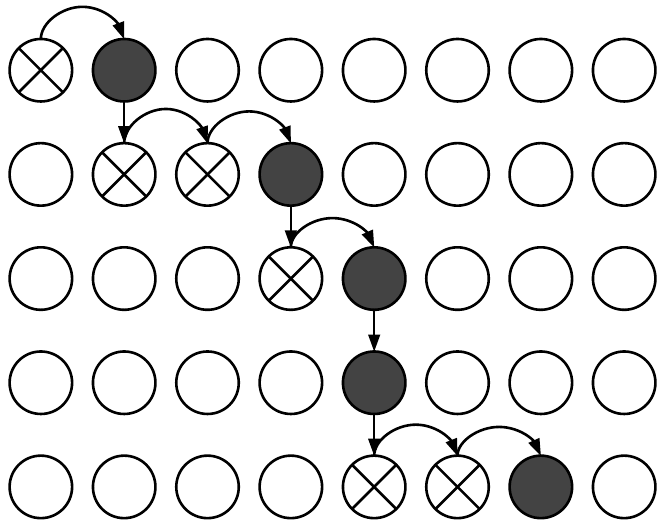}
        \caption{Monotonic attention.}
        \label{fig:monotonic}
    \end{subfigure}
    \begin{subfigure}[b]{0.33\textwidth}
        \includegraphics[width=\textwidth]{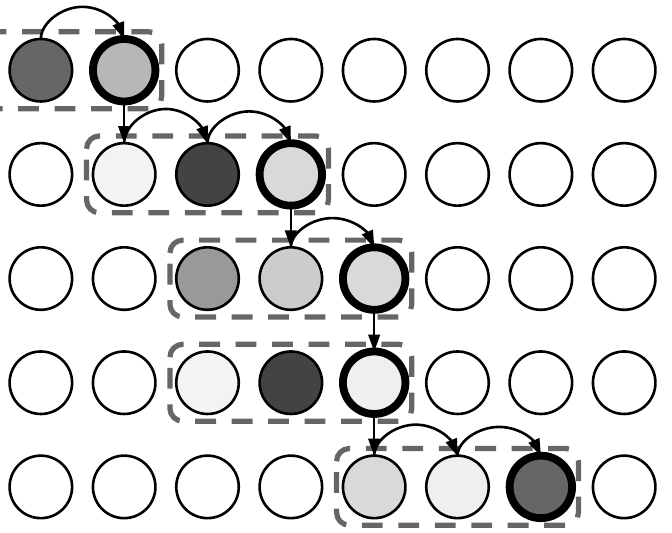}
        \caption{MoChA attention.}
        \label{fig:mocha}
    \end{subfigure}
    \begin{subfigure}[b]{0.33\textwidth}
        \includegraphics[width=\textwidth]{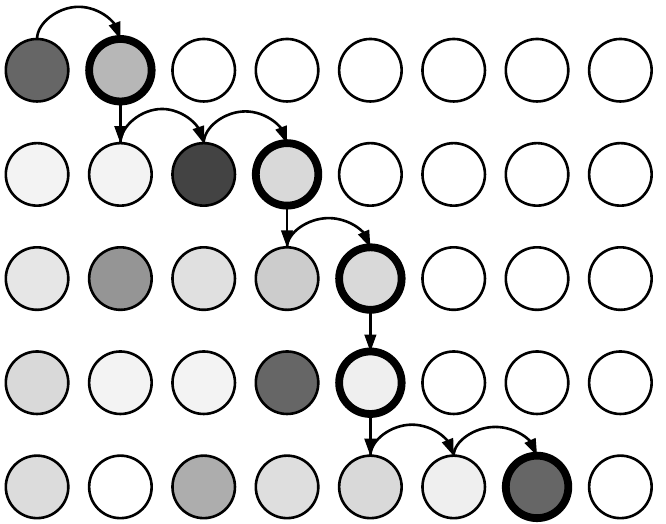}
        \caption{MILk attention.}
        \label{fig:milk}
    \end{subfigure}
    \begin{subfigure}[b]{0.33\textwidth}
        \includegraphics[width=\textwidth]{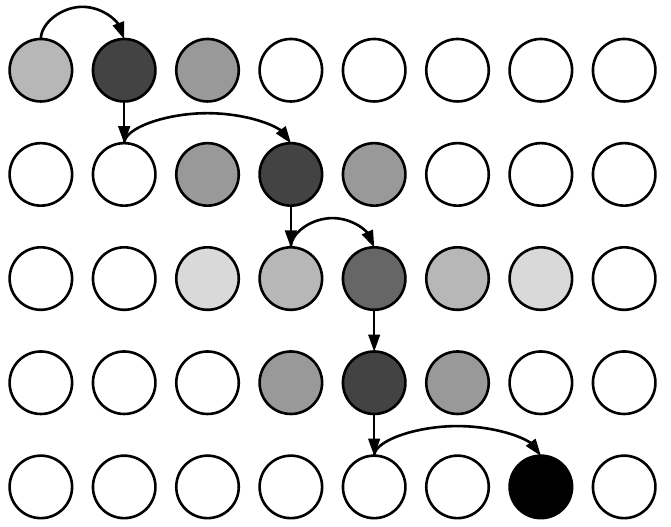}
        \caption{GMM monotonic attention.}
        \label{fig:gmm}
    \end{subfigure}
    \begin{subfigure}[b]{0.33\textwidth}
        \includegraphics[width=\textwidth]{figs/monotonic.pdf}
        \caption{RNN-T.}
        \label{fig:rnnt}
    \end{subfigure}
    \caption{A simple diagram comparing the end-to-end approaches evaluated in this work.  The horizontal axis corresponds to the encoder steps while the vertical axis corresponds to prediction steps.  (a)-(e) are attention-based models.  The GMM monotonic attention use mixture of multiple distribution, but in (e) for the clarity of the illustration we plot a single distribution case.  The monotonic attention (b) and RNN-T (f) exhibit the same behavior in terms of selecting encoder hidden states for making prediction, but differ in how the selected encoder state being used by the decoder.}
    \label{fig:models}
\end{figure*}

A popular and effective approach to building end-to-end models is with attention-based models.  The common architecture of attention-based models consist of an encoder, a decoder, and an attention mechanism:
\begin{align}
    h_j =& [EncoderForwardRNN(x_j, h_{j - 1}), \nonumber\\
         &  EncoderBackwardRNN(x_j, h_{j + 1})] \\
    s_i =& \decoder(y_{i - 1}, s_{i - 1}, c_i) \\
    y_i =& \outpt(s_i, c_i)
\end{align}
where $h_j$ is the encoder state at input timestep $j$, $s_i$ is the decoder state at output timestep $i$, and $c_i$ is a context vector.  The models explored in this work use encoders with bi-directional RNNs.  The context vector is computed based on the encoder hidden states through the use of an attention mechanism.  Within this class of models, a variety of underlying attention mechanisms can be used.  Below we describe the computation of $c_i$ with respect to each attention mechanism explored in this work.

\subsection{Soft attention}

In the standard soft attention model, attention context is computed based on the entire sequence of encoder hidden states, which fundamentally limits the length of sequences this attention model can scale to, for two reasons. Firstly, attention computation cost is linear in the sequence length. When the source sequence is very long, the cost of computing the attention context is too high for each decoding step. Secondly, when sequence is very long, attention mechanism can easily get confused, resulting in non-monotonically moving attention head. In our experiments, we show that soft attention model trained on short utterances has difficulty scaling to long utterances and suffers from a high deletion rate.

This problem with soft attention model can be mitigated by exploiting the fact that in ASR, alignment between source and target is always monotonic. Based on where the attention head was at the previous decoding step, in computing the attention context for the next decoding step, one can limit focus to only a subsequence of the encoder hidden states. In the rest of this section, we describe a few variants of this soft attention model that exploit this spatial constraint in different ways.

\subsection{Monotonic attention}

\cite{Raffel2017} proposed an attention mechanism that scans the sequence of the encoder hidden states in a left-to-right order and selects a particular encoder state for computing the context vector.  This selection probability is computed through the use of an energy function that is passed through a logistic function to parameterize a Bernoulli random variable.
The hard monotonic decision process however prevents the attention mechanism from being trained with standard backpropagation.
To solve this problem, \cite{Raffel2017} proposed to replace this one-hot attention vector (it is $1.0$ for the chosen encoder state, and $0.0$ elsewhere) with a soft expected attention probability vector during training.

With the monotonic attention mechanism, at each decoding step the decision process starts from the previously selected state and makes a frame-by-frame decision sequentially.  This focuses the attention decision to only a sub-sequence of the encoder output, and thus in theory has better potential to scale to long-form utterances compared to the standard soft attention mechanism.

\subsection{Monotonic Chunkwise Attention}

While the monotonic attention mechanism provides better scalability for the long sequences, it limits itself to consider only a single step of the encoder states and therefore reduces the power of the attention model.  The monotonic chunkwise attention (MoChA)~\cite{Chiu2018} mechanism remedies this by allowing an additional lookback window to apply soft attention.
The context vector in MoChA is more similar to the standard soft attention which contains weighted combination of a set of encoder states, as opposed to the monotonic attention mechanism which uses only a single step's encoder state.

\subsection{Monotonic Infinite Lookback Attention}


The MoChA mechanism extends the capability of the monotonic attention mechanism by allowing it to look back a fixed window of encoder states from the current attention head. This fixed window size may still limit the full potential of the attention mechanism. The monotonic infinite lookback attention (MILK) mechanism was proposed in ~\cite{Arivazhagan2019} to allow the attention window to look back all the way to the beginning of the sequence. 


The MILK attention mechanism has to be coupled with a latency loss that encourages the model to make the emission decision earlier. To see why, without the latency loss, the model may decide to wait until the end of source sequence to make even the first prediction, which then effectively recovers the standard soft attention mechanism and loses the benefit brought by the monotonic attention mechanism. 

\subsection{GMM monotonic attention}
\cite{Graves2013generating} proposed GMM attention to explicitly enforce the mode of probability mass generated by the current attention modules that are always moving incrementally to the end of the source sequence. The
selection probability into $h_j$ at timestep $i$ ($1 \leq i \leq T$) is defined by the following a mixture of $K$ Gaussian functions:
\begin{align}
    \alpha_{i, j} &= \sum_{k=1}^K w^k_j \frac{1}{\sqrt{2\pi v^k_j + \mathrm{vfloor}}}\exp(-\frac{(j-\mu^k_j)^2}{2v_j^k + \mathrm{vfloor}})\label{eq:gmm_alpha}
\end{align}
where
\begin{align}
(\gamma^k_j, \beta^k_j, \kappa^k_j) &= \mathrm{FeedForwardNet}(s_i) \nonumber \\
w^k_j &= \mathrm{Softmax}(\gamma^k_j) \nonumber\\
v^k_j &= \exp(\beta^k_j) \nonumber\\
\mu^k_j &= \exp(\kappa^k_j) + \mu^k_{j-1} \label{eq:gmm_params}
\end{align}
The parameters of GMM (Eq.~\ref{eq:gmm_params}) distribution are estimated by a single layer feedforward network. We added a variance floor ($\mathrm{vfloor}=1\mathrm{e}{-8}$ to make training more stable.

\section{RNN Transducer}
\label{sec:rnnt}
Besides attention-based models, RNN-T~\cite{Graves12,Graves2013} has shown successful results on building end-to-end models for speech recognition~\cite{He2019}.  RNN-T is most similar to the monotonic attention model in that both models scan the encoder states sequentially to select a particular encoder state as the next context vector.  This sequential scanning property is essential in allowing RNN-T to scale well to long utterances.  

At decoding time, given a new encoder state, both the RNN-T model and the monotonic attention model make a ``predict/no-predict'' decision. The two models however differ in how the ``predict/no-predict'' decision affects decoder's token prediction.  In the monotonic attention mechanism, if a ``predict'' decision was made, the decoder then takes the encoder state as attention context to make a token prediction.  If a ``no-predict'' decision was made instead, then the decoder does nothing, and simply waits for the next encoder state.  In a contrary, RNN-T takes ``no-predict'' as one of the output tokens.  Essentially in RNN-T ``predict/no-predict'' decision happens at the output level. 

In training the RNN-T model, we compute the sum of probabilities over all valid combinations of ``predict/no-predict'' choices with an efficient dynamic programming algorithm, see ~\cite{Graves12,Graves2013} for details. In training the monotonic attention model, we compute the expected attention probabilities over the source sequence in order to avoid backpropagating through discrete ``predict/no-predict'' choices, see \cite{Raffel2017} for more details.

The comparison of each model's mechanism on selecting encoder state for predictions are shown in Fig~\ref{fig:models}.

\section{OVERLAPPING INFERENCE}
\label{sec:overlap}
\begin{figure*}[t]
  \centering
  \includegraphics[width=0.8\textwidth]{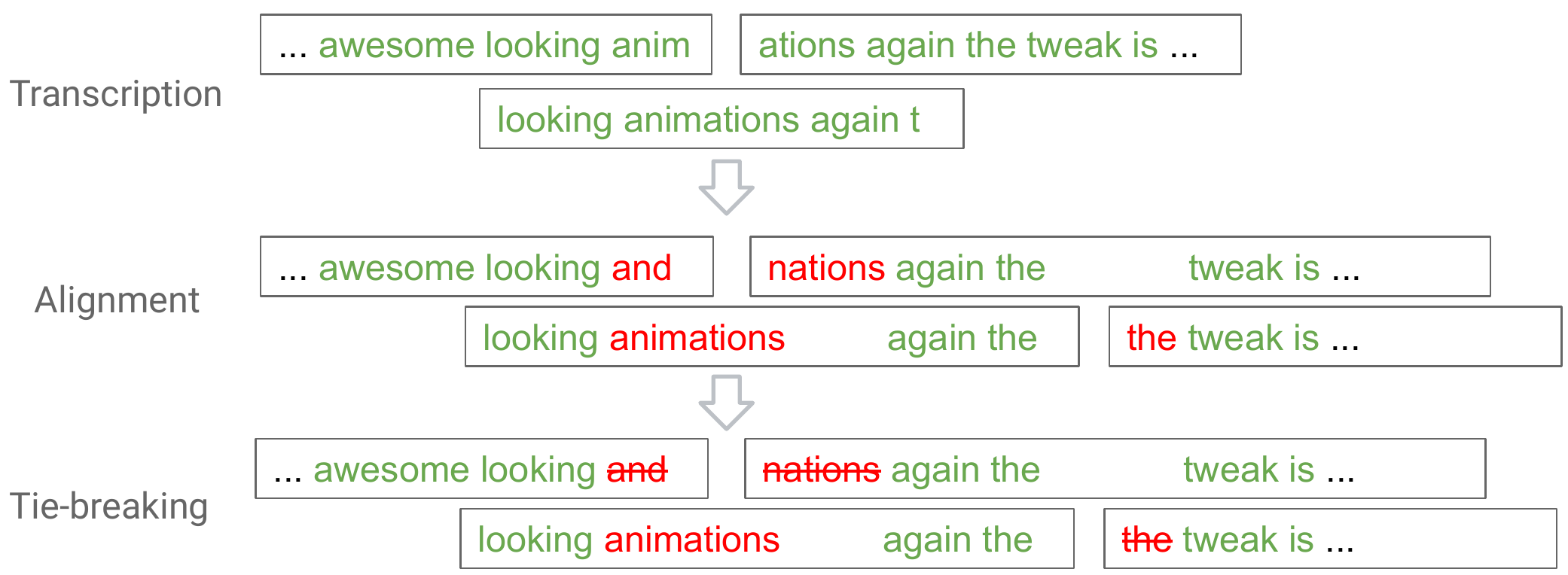}
  \caption{Overlapping inference.  The algorithm
  first breaks a long utterance into overlapped segments, each of which is then transcribed independently. It then merges the transcripts from overlapped segments into a consensus transcript for the original long utterance. In case there are conflicts in predictions, it prefers the predictions further from the utterance boundary.}
  \label{fig:overlap}
\end{figure*}

Overlapping inference is a decoding strategy that we proposed in order to further improve attention based model performance on long form audios.  In general, due to various constraints, training of the end-to-end models is often on short utterances only.  Hence, there is an inherent train and inference mismatch when a model trained on short utterances alone is used to transcribe long utterances.  Overlapping inference is designed to bridge this train/inference mismatch.

A straightforward approach to this train/inference mismatch problem is to break a long utterance into fixed length segments, and then transcribe each segment independently.  This however will result in deteriorated performance especially at the segment boundaries, for two reasons. First, a segment boundary may cut through the middle of a word, making it impossible to recover the original word from either of the segments, as illustrated in Fig~\ref{fig:overlap}.  Second, the recognition quality can be poor at the beginning of a segment due to lack of context.  A smarter segmenter can be used, e.g. based on some voice activity detection algorithms to segment only when there is a sufficiently long silence.  However, those segmenters can still produce long segments when no sufficiently long pause/silence is detected.

Overlapping inference improves over the aforementioned fixed-length segmenter or smarter segmenter based approaches, with a simple trick: it breaks a long utterance into overlapping segments.  In our experiments, we chose $50\%$ overlap, which means that any point of audio is covered by exactly two segments.  The information loss at a boundary of a segment can always be recovered by referencing to the other overlapping segment.

\subsection{Combine overlapping windows with 50\% overlap}
\label{sec:overlapping_no_timing}

In overlapping inference, we create windows of a fixed length $L$ and fixed overlap $D=L/2$.  A special property of this setup is any word in the utterance will always get recognized twice by two consecutive windows. Conveniently, two parallel hypotheses can be constructed as the concatenation of the odd numbered windows and the even numbered windows, respectively:

\begin{align*}
Y^{o} & =\ldots,y_{1}^{2k},\ldots,y_{r_{2k}}^{2k},y_{1}^{2k+2},\ldots,y_{r_{2k+2}}^{2k+2},\ldots\\
Y^{e} & =\ldots,y_{1}^{2k-1},\ldots,y_{r_{2k-1}}^{2k-1},y_{1}^{2k+1},\ldots,y_{r_{2k+1}}^{2k+1},\ldots,\\
\end{align*}
where $y_{i}^{j}$ denotes the $i_{th}$ recognized word in the $j_{th}$ window.

\subsubsection{Matching $Y^o$ and $Y^e$}

Next, we search for the best matching between $Y^o$ and $Y^e$.  The problem closely resembles the editing distance minimization that is commonly used in WER calculation, with only a minor difference in that it constrains to disallow words that are more than one window away from being matched.  In other words, the process only matches words where their windows overlap.  The solution can be found efficiently using a dynamic programming algorithm.  The result of matching is a sequence of word pairs $\langle p_1, q_1 \rangle, \langle p_2, q_2 \rangle,\ldots,\langle p_T, q_T \rangle$, where
\begin{align*}
\langle p_i, q_i \rangle & = \begin{cases}
\langle y^o_{i_o, j_o}, y^e_{i_e, j_e}\rangle & \text{if } y^o_{i_o, j_o} \text{ aligned with } y^e_{i_e, j_e}\\
\langle \varnothing, y^e_{i_e, j_e}\rangle & \text{if } y^e_{i_e, j_e} \text{ has no alignment}\\
\langle y^o_{i_o, j_o}, \varnothing\rangle & \text{if } y^o_{i_o, j_o} \text{ has no alignment,}\\
\end{cases}
\end{align*}
$i$ is the pair index, $T$ is the total number of matched pairs, $y^o_{i_o, j_o}$ denotes $i_{o}th$ word at $j_{o}th$ window from $Y^o$, $y^e_{i_e, j_e}$ denotes $i_{e}th$ word at $j_{e}th$ window from $Y^e$, $\varnothing$ denotes no predictions.

\subsubsection{Tie-breaking}

During inference models generally see more contextual information for words further away from window boundaries, and therefore our approach assign higher confidence for those words.  Concretely, we define a confidence score based on the relative location of a word in the window:
\begin{align*}
f(y^i_j, S^i, L)=-|s^i_j - (S^i+L/2)|
\label{eq:conf_score_v1}
\end{align*}
where $S^i$ is the starting time of the $i_{th}$ window and $s^i_j$ is the starting time of the word $j$ at window $i$.  The score peaks at the center of the window and linearly decays towards boundaries on both sides.  For the RNN-T model we define $s^i_j$ as the time step that the model decides to emit the word, and in the case of no prediction the process use the starting time of the matched word as the starting time of $\varnothing$.  For attention-based models, we use the relative position of the word and simplify the equation to
\begin{align*}
f(y^i_j, S^i, L)=-|j/C_i - 1/2|
\end{align*}
where $C_i$ denotes the number of matched words in window $i$.  The final hypothesis selects words with higher confidence score:

\begin{align*}
Y^{*}_i & =\begin{cases}
p_i & \text{if }f(p_i)\geq f(q_i)\\
q_i & \text{if }f(p_i)<f(q_i)
\end{cases}
\end{align*}

We note that with $50\%$ overlap, overlapping inference increases inference computation cost to 2x.  We are exploring ways to cut down this computational cost by reducing the overlap at the boundaries.

\section{EXPERIMENTS}
\label{sec:exp}
\begin{table*}[t]
  \begin{center}
  \begin{tabular}{c|rrrrr}
    Model & Original & Segment $16s$ & Segment $30s$ & Overlapping inference $8s$ & Overlapping inference $16s$ \\ \hline
    Soft & $67.1$ & $13.3$ & $17.1$ & $11.7$ & $12.0$ \\
    Monotonic & $55.6$ & $15.5$ & $16.4$ & $12.8$ & $13.6$ \\
    MoChA & $61.8$ & $14.4$ & $17.0$ & $12.6$ & $12.9$ \\
    MILK & $49.4$ & $\mathbf{13.0}$ & $17.3$ & $\mathbf{11.6}$ & $\mathbf{11.7}$ \\
    GMM & $18.8$ & $13.6$ & $14.3$ & $11.8$ & $12.5$ \\
    RNN-T & $\mathbf{12.0}$ & $13.2$ & $13.1$ & $12.3$ & $11.9$ \\ \hline
    CTC $5\times600$* & $14.5$ & & & & \\
    CTC $7\times1000$* & $13.5$ & & & &
  \end{tabular}
  \end{center}
  \caption{Word-error-rates of end-to-end models on YouTube test set.  For the \textit{Original}, the utterances are segmented based on the appearance of silence.  The resulting segmented utterances ranges from a few seconds to a few minutes long.  \textit{Segment $16s$} corresponds to imposing an additional segmentation threshold on the silence-segmented utterances where utterances longer than $16$ seconds are force segmented.  We compare the results of using the overlapping inference with chunk size $8$ and $16$ seconds.  *: the word-error-rates of CTC models are from~\cite{Soltau2017} and are with language models.
}\label{table:exp}
\end{table*}



We conduct our experiments on the Youtube data set, the same data set as used in~\cite{Soltau2017}. YouTube videos cover a wide variety of different domains~\cite{Narayanan2018}, and have a wide range of length distributions, making it an ideal test bed for this long-form transcription study.

Same as in ~\cite{Soltau2017}, our training data consists of english utterances extracted according to an island of confidence approach~\cite{Liao2013}. To run the algorithm, a pre-existing ASR model is used, which is a conventional ASR system. The pre-existing ASR model is being adapted on a per-video basis with a per-video specific language model built using the user uploaded transcripts. The segments (also called `island') where the ASR produced transcript matches the user uploaded transcript exactly are being extracted as our training data. In total, there were 125 thousand hours of data being extracted. Most of the extracted segments are short utterances, with $99-th$ percentile at $16.74$ seconds. At training time, we cap our training utterances to be at most $17.28$ seconds long. 

The test set is comprised of 296 videos with length ranging from $2$ to $9$ minutes. The total duration of the test videos is $26$ hours. The videos in the test set are much longer than the training samples, and hence ASR models trained on short utterances need to be able to scale to long videos to be able to perform well.

The input uses $80$-dimensional log-Mel features, computed with a $25$ms window and shifted every $10$ms.  Each input time step stacks $5$ frames of these features, with $2$ frames from the left and $2$ frames from the right, and downsampled to a $30$ms frame rate.  We compared the soft attention, monotonic attention, monotonic chunkwise attention, monotonic infinite lookback attention, GMM-based monotonic attention, and RNN-T models.  All our end-to-end models have an encoder composed of $5$ layers of bi-directional LSTMs with dimension $1024$ ($512$ each direction).  The architecture of attention models follow the same design of the bi-directional model as described in~\cite{Chiu2018state}, but do not use scheduled sampling, minimum word error rate training~\cite{RohitSeq17}, and the second-pass language model rescoring.  For the MoChA model we use chunk size of $8$.  In the MILK model we applied a latency loss different from the the one proposed in \cite{Arivazhagan2019}, as the original latency loss is tailored for machine translation where the source and target sequence have similar length.  Our latency loss minimize the root-mean-square value of the interval between two consecutive emissions:
\begin{align}
  Latency &= \sum_{j=1}^{|x|}[\alpha_{i,j} \sum_{k=1}^{|x|}\alpha_{i-1,k} Delay(j - k)]^2 \\
  Delay(x) &= \left\{
    \begin{array}{ll}
      x & x > 0 \\
      0 & x <= 0
    \end{array}\right.
\end{align}
In the GMM monotonic attention model, there are $5$ mixture components.  As of the RNN-T model, while the encoder is the same as attention-based models, the prediction network has $2$ LSTM layers with $2048$ hidden units and a $640$-dimensional projection~\cite{Pang2018} per layer.  The output network has $640$ hidden units and the softmax layer predicts graphemes which has $76$ units.  The prediction and output network architecture is the same as the RNN-T grapheme model described in~\cite{He2019}.  All models are implemented with Tensorflow-Lingvo \cite{lingvo}, and the RNN-T model further utilize techniques described in \cite{Sim17,Bagby2018} to improve training efficiency.

The results are summarized in Table~\ref{table:exp}.  As shown from the word-error-rates (WERs), attention models have problems scaling up to long utterances. In particular, those high WERs are due to the high deletion errors. For example, of the 67.1 WERs of the soft attention model, it consists of 63.2 deletion errors, 0.7 insertion errors, and 3.1 substitution error.  This implies that on recognizing long-form utterances, attention models failed to generalize to the whole utterance and only produce transcripts for a sub-sequence of the utterance. The attention approaches that utilize monotonic alignment property all perform better than the vanilla soft-attention model, but they still exhibit serious problems generalizing to long utterances.
Among all the attention mechanisms, the GMM-based monotonic attention model perform the best. The RNN-T model is robust to long-form utterances, and in fact achieves the best quality when longer utterances are being preserved.  In \cite{Soltau2017} they reported the CTC end-to-end models outperform phone-based models, and achieve $13.5$ WER with $7\times1000$ bi-directional LSTMS.  The attention models are significantly worse compared to the CTC models on long-form speech recognition, but the RNN-T model is able to outperform them.

The issue of long-form speech recognition with attention models can be addressed with the use of overlapping inference.  Through segmenting the utterances into smaller chunks that are closer to the training utterances' length, the attention-based model can provide competitive quality.  Simply segmenting utterances into smaller sub-sequences can lead to missing context at the segmentation boundaries, which explains the quality loss between the "Segment 16s" approach and the "Overlapping inference 16s" approach.  With the help of the overlapping inference, the MILK model provides the best quality compared to other end-to-end models, though the delta between other attention based models aren't large.

When segmenting utterances into shorter sub-sequences, the model also loses contextual information.  The results of the RNN-T model provides a reference for measuring this information loss.  On regular segmentation with $16$ seconds threshold the RNN-T model observe a $10\%$ relative quality loss.  In comparison the overlapping inference with $16$ seconds segmentation was able to amend this quality loss.

\section{CONCLUSIONS}
\label{ssec:conclusions}
In this work we compare various end-to-end models for long-form speech recognition. The end-to-end models are trained on short utterances and evaluated on much longer utterances. The evaluation results show that the RNN-T model is able to scale to recognize long utterances and provides very strong quality. The attention-based models in general can't generalize to long-form utterances. The GMM-based monotonic attention model performs the best among all attention model on this task, but still significantly lags behind the RNN-T model. We show that by incorporating overlapping inference, we can improve the performance of the attention-based models to be very competitive to that of the RNN-T model.

\section{ACKNOWLEDGEMENT}
\label{ssec:ack}
We would like to thank Hagen Soltau's contributions on the YouTube dataset and the recipe for training the RNN-T model.

\bibliographystyle{IEEEbib}
\bibliography{ref}

\end{document}